# روش جدیدی برای نهان‌نگاری تطبیقی در تصاویر

کاظم غضنفری[1]، رضا صفابخش[2]
[1]دانشکده مهندسی کامپیوتر، دانشگاه صنعتی امیرکبیر، kazemmit@aut.ac.ir
[2]دانشکده مهندسی کامپیوتر، دانشگاه صنعتی امیرکبیر، safa@aut.ac.ir

**چکیده** – هدف از نهان‌نگاری ایجاد یک کانال محرمانه و امن بین فرستنده و گیرنده است. یکی از رایج‌ترین و ساده‌ترین روش‌های ایجاد این کانال، درج اطلاعات محرمانه در بیت کم‌ارزش مقادیر پیکسل‌های تصویر می‌باشد. چون این روش مشخصات آماری و ادراکی تصویر را به مقدار زیادی تغییر می‌دهد، مورد توجه حملات گوناگونی قرار گرفته است. یکی از این حملات، حمله آماری مراتب بالا می‌باشد. این حمله همبستگی پیکسل‌ها را نسبت به هم در نظر می‌گیرد و با توجه به تغییرات بوجود آمده ناشی از درج اطلاعات محرمانه اقدام به نهان‌کاوی می‌کند. بنابراین اگر بتوان تغییر مقادیر پیکسل‌ها را که ناشی از درج اطلاعات محرمانه است، کاهش داد، می‌توان تا حدود زیادی تغییرات آماری مراتب بالای بوجود آمده ناشی از درج را کاهش داد. برای این منظور در این مقاله تکنیک جدیدی برای نهان‌نگاری تطبیقی ارائه می‌شود بطوریکه منجر به کمترین تغییرات در نواحی مسطح تصویر می‌شود. به کمک این تکنیک می‌توان با اعمال تغییر کوچک در تعداد کمی از پیکسل‌های تصویر، اطلاعات محرمانه را در تصویر درج نمود. همچنین در این مقاله روش جدیدی برای درج پیام در نواحی نویزی تصویر ارائه می‌شود که منجر به افزایش ظرفیت تا 1/5 برابر در این نواحی می‌شود. نتایج آزمایشات نشان می‌دهد که روش پیشنهادی با ظرفیت درج یکسان در مقایسه با تعدادی از روش‌های نهان‌نگاری در حوزه مکان، منجر به خرابی‌های ادراکی کمتری شده و حملات هیستوگرام و آماری مراتب بالا با قدرت بسیار کمتری قادر به تشخیص آن می‌باشند.

**کلید واژه**— نهان‌نگاری، نهان‌کاوی، حمله هیستوگرام و حمله آماری مراتب بالا، درج در بیت کم ارزش، ماتریس خودهمبستگی

## 1- مقدمه

با توجه به رشد سریع ارتباطات و اینترنت، انتقال داده‌ها در قالب دیجیتال امری مرسوم شده است. یک مسئله مهم در رابطه با انتقال اطلاعات دیجیتال از طریق اینترنت آن است که امنیت انتقال اطلاعات تضمین شود. حال هدف از نهان‌نگاری، ارسال پیام‌های محرمانه بصورت مخفیانه روی شبکه می‌باشد، بطوریکه مهاجمان متوجه انتقال داده‌های مخفی و امنیتی در کانال ارتباطی نشوند[1،2،3].

روش‌های گوناگونی برای نهان‌نگاری اطلاعات در تصویر ارائه شده است. در یک دسته‌بندی، روش‌های نهان‌نگاری را بر اساس حوزه درج به دو دسته اصلی تقسیم بندی می‌کنند: نهان‌نگاری در حوزه مکان و نهان‌نگاری در حوزه تبدیل. روش‌های نهان‌نگاری در حوزه مکان، اطلاعات محرمانه را بطور مستقیم در مقدار پیکسل‌ها درج می‌کنند، درحالیکه روش‌های نهان‌نگاری در حوزه تبدیل ابتدا تصویر را به کمک یکی از تبدیلات مانند $DCT$، $DWT$ و $DFT$ به حوزه فرکانس انتقال می‌دهند، سپس عمل نهان‌نگاری را در این حوزه انجام می‌دهند. درج اطلاعات در تصویر منجر به تغییرات گوناگونی می‌شود که به سه دسته کلی تقسیم می‌شوند: تغییرات ادراکی، آماری و ساختاری، که این تغییرات زمینه تشخیص وجود پیام محرمانه در تصویر را برای مهاجمان فراهم می‌نماید[1،2،3].

حمله آماری مراتب بالا یکی از رایج‌ترین حملات می‌باشد که از تغییرات آماری بوجود آمده بر اثر درج اطلاعات در تصویر برای تحلیل استفاده می‌کند. بر اساس اینکه عمل نهان‌نگاری در چه حوزه‌ای انجام شده است، این نوع حمله قابل انجام می‌باشد و ما نیازمند روش‌هایی هستیم که اطلاعات آماری مراتب بالای تصویر (یا ضرایب تبدیل) را پس از درج در حوزه مکان (یا تبدیل) حفظ کنند. در این مقاله ما روش جدیدی برای مقابله با حملاتی که از اطلاعات آماری مرتبه بالا استفاده می‌کنند ارائه خواهیم داد.

## 2- اطلاعات آماری تصویر

بطور کلی اطلاعات آماری قابل استخراج از تصویر را به دو دسته اصلی تقسیم می‌کنند که شامل اطلاعات آماری مرتبه اول و اطلاعات آماری مراتب بالاتر می‌باشد. یکی از مهمترین اطلاعات آماری مرتبه اول، هیستوگرام تصویر می‌باشد. با توجه به اینکه انجام عمل نهان‌نگاری باعث بوجود آمدن تغییرات قابل توجهی در هیستوگرام تصویر می‌شود، روش‌های زیادی از این تغییرات استفاده کرده و

اقدام به نهان‌کاوی کرده‌اند[4،5،6]. برای مقابله با این حمله روش‌های زیادی ارائه شده است[7،8،9،10].

برای نمونه در روش فرنز[7] بر اساس اینکه از چند بیت کم ارزش برای درج استفاده خواهد شد، تعدادی گروه (واحد) در هیستوگرام تصویر تشکیل می‌شود. بعد از مشخص شدن این گروه‌ها، توزیع هر کدام از این گروه‌ها بدست آمده و سپس پیام به قطعات کوچکی تقسیم می‌شود و بر اساس توزیعی که هر قطعه پیام دارد، گروه مناسب برای درج آن قطعه انتخاب شده و عمل درج در آن انجام می‌شود.

در روش شارپ[8]، به جای اینکه عمل درج بیت پیام بطور مستقیم در بیت کم‌ارزش انجام شود، اگر بیت پیام با بیت‌کم‌ارزش برابر باشد، تغییری اعمال نمی‌شود، ولی اگر برابر نباشد، بر اساس مقدار احتمالی که بصورت پارامتر مشخص می‌شود، بیت‌کم‌ارزش را یک واحد کم یا اضافه می‌کنند. همچنین روشی در [9] ارائه شده است که در آن، در صورتیکه نیاز به تغییر بیت‌کم‌ارزش باشد، از طریق جابه‌جایی ضرایب در عوض تغییر آنها، هیستوگرام تصویر یا ضرایب تبدیل مربوطه حفظ می‌شود.

در روش وو و همکاران[10]، سعی می‌شود از طریق درج عمدی تعدادی صفر یا یک، هیستوگرام تصویر حفظ شود. در واقع در این روش از طریق قربانی کردن تعدادی پیکسل، هیستوگرام تصویر حفظ می‌شود.

چون هیستوگرام تصویر وابستگی بین پیکسل‌ها را در نظر نمی‌گیرد، این ویژگی جزء اطلاعات آماری مرتبه اول در نظر گرفته می‌شود. اطلاعات آماری مراتب بالاتر، به اطلاعاتی گفته می‌شود که وابستگی پیکسل‌ها را نسبت به یکدیگر در نظر گیرد. یکی از مهمترین اطلاعات آماری مراتب بالا که در نهان‌کاوی مورد استفاده قرار می‌گیرد، ماتریس خودهمبستگی می‌باشد. به دلیل اینکه انجام عمل نهان‌نگاری باعث می‌شود این اطلاعات تغییر پیدا کند، روش‌های زیادی از این ماتریس بهره گرفته و اقدام به نهان‌کاوی نموده‌اند[7،11،12].

یک روش برای توصیف همبستگی بین پیکسل‌ها استفاده از ماتریس‌های خودهمبستگی است. ماتریس خودهمبستگی، یک ماتریس دو بعدی است که مدخل‌های آن بیانگر فرکانس رخ دادن دو سطح خاکستری است که به فاصله و در جهت خاصی قرار دارند. زوج $(\Delta x, \Delta y)$ بیانگر فاصله و جهت می‌باشند. این ماتریس به نوعی بیانگر اطلاعات آماری مرتبه دو می‌باشد، چون همبستگی پیکسل‌ها را دوبه‌دو در نظر گرفته است. برای سطح خاکستری $g(x,y)$ مربوط به پیکسل در مکان $(x,y)$ مدخل $c_{i,j}$ ماتریس خودهمبستگی، فرکانس تعداد زوج‌هایی را بیان می‌کند که در رابطه زیر صدق کنند.

$$(g(x,y) = i) \wedge (g(x+\Delta x, y+\Delta y) = j) \quad (1)$$

در اینصورت برای هر زوج $(\Delta x, \Delta y)$ یک عدد ماتریس خودهمبستگی ایجاد می‌شود. توصیف تاثیر درج روی این ماتریس نسبت به هیستوگرام تصویر بسیار سخت‌تر است، چون در اینجا ارتباط پیکسل‌ها دوبه‌دو در نظر گرفته می‌شود. در تصاویر طبیعی سطح خاکستری پیکسل‌های مجاور پیوسته و نرم می‌باشد، به عبارت دیگر، همبستگی موجود بین پیکسل‌های مجاور نسبت به همبستگی موجود بین پیکسل‌هایی که دارای فاصله زیادی از هم می‌باشند، بسیار بیشتر است. به همین دلیل در روش‌های نهان‌کاوی معمولاً ماتریس‌های خودهمبستگی را فقط برای پیکسل‌های همسایه محاسبه می‌کنند. به عبارت دیگر، ماتریس‌های همبستگی را به ازای زوج‌های زیر محاسبه می‌کنند.

$$(\Delta x, \Delta y) \in \{(1,0), (-1,1), (0,1), (1,1),$$
$$(-1,-1), (0,-1), (1,-1), (-1,0)\} \quad (2)$$

البته در مقالات گوناگون، تمام همسایگی‌های فوق مورد استفاده قرار نگرفته است. بطور مثال در مرجع [11] زوج‌های:

$$(\Delta x, \Delta y) \in \{(-1,1), (0,1), (-1,-1), (-1,0)\}$$

و در مرجع [12] زوج‌های:

$$(\Delta x, \Delta y) \in \{(1,0), (0,1)\}$$

و همچنین در مرجع [7] زوج‌های زیر مورد استفاده قرار گرفته است:

$$(\Delta x, \Delta y) \in \{(1,0), (-1,1), (0,1), (1,1)\}$$

با توجه به اینکه در تصاویر سطح خاکستری برای نمایش مقدار پیکسل‌ها از 8 بیت استفاده می‌شود، ابعاد این ماتریس برابر 256×256 خواهد بود. البته در مقاله [12] برای کاهش حجم محاسبات و ابعاد ماتریس ابتدا مشتق تصویر محاسبه شده، بعد از یک مرحله آستانه‌سازی (شبه آستانه‌سازی) مقادیر تصاویر مشتق، اقدام به محاسبه این ماتریس برای تصویر مشتق آسانه‌ای شده کرده است. واضح است در اینصورت ابعاد ماتریس به شدت کاهش پیدا می‌کند.

"تنها آن سطوح خاکستری می‌توانند در فرآیند درج مورد استفاده قرار گیرند که نسبت به سطح خاکستری جدید حاصل از درج مستقل تصادفی باشند."

همانطور که از مطلب فوق مشخص است، این روش نهان‌نگاری در تصویر را فقط در نواحی نویزی امکان‌پذیر می‌داند که این عامل باعث می‌شود ظرفیت نهان‌نگاری به شدت برای تصاویری که نویزی نیستند کاهش پیدا کند. برای این منظور در این مقاله روش جدیدی برای نهان‌نگاری اطلاعات در تصاویر بدون نویز ارائه می‌شود که بدون کاهش چشمگیر ظرفیت، از تغییرات آماری مرتبه دو جلوگیری می‌کند. در بخش بعدی جزئیات مربوط به روش پیشنهادی ارائه می‌شود.

## 3- روش پیشنهادی

در روش پیشنهادی، فرآیند نهان‌نگاری به دو مرحله تقسیم می‌شود. در مرحله اول تصویر به بلوک‌هایی با ابعاد n×n تقسیم می‌شود. این بلوک‌ها به دو دسته اصلی بلوک‌های نویزی و بلوک‌های مسطح تقسیم‌بندی می‌شوند. در مرحله بعد، بر اساس اینکه بلوک مورد نظر مسطح یا نویزی است، از روشی خاص استفاده شده و پیام مورد نظر در آن بلوک درج می‌شود. در یکی از دسته‌بندی‌های روش‌های نهان‌نگاری، تکنیک‌های نهان‌نگاری را بر اساس لزوم اینکه تصویر پوششی در مقصد موجود باشد یا نباشد (برای استخراج پیام)دسته‌بندی می‌کنند. در روشی که در این مقاله ارائه می‌شود، گیرنده برای استخراج پیام باید تصویر پوششی را در اختیار داشته باشد.

### 3-1- ناحیه‌بندی تصویر

روش‌های مختلفی برای تشخیص اینکه بخشی از تصویر نویزی هست یا نیست موجود است. با توجه به اینکه اطلاعات آماری مراتب بالا بشدت به نواحی مسطح حساس می‌باشد، می‌توان نتیجه گرفت که آن نواحی که مسطح نیستند را می‌توان نواحی نویزی تلقی کرد. تغییر در سطح خاکستری پیکسل‌های آن نواحی تاثیر کمی روی نواحی نزدیک قطر اصلی ماتریس خودهمبستگی می‌گذارد. برای این منظور فرض کنید تصویر به بلوک‌هایی با اندازه n×n تقسیم شده است. بلوک B را بلوک مسطح گویند اگر تابع آستانه‌ساز Threshold(d) مقدار یک برای آن تولید کند.

$$Threshold(d) = \begin{cases} 1 & d < T \\ 0 & d \geq T \end{cases} \quad (3)$$

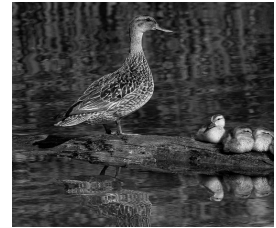

(الف)

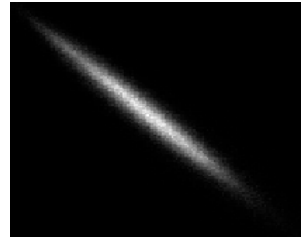 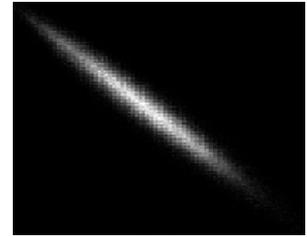

(ب)        (ج)

شکل شماره 1: (الف) تصویر مورد استفاده. (ب) ماتریس خودهمبستگی قبل از درج پیام. (ج) ماتریس خود همبستگی بعد از درج پیام

همانطور که از تعریف ماتریس خودهمبستگی مشخص است، با توجه به اینکه پیکسل‌های مجاور دارای همبستگی زیادی می‌باشند و از طرف دیگر تغییرات سطوح خاکستری در تصاویر نرم و پیوسته است، مقادیر خانه‌های نزدیک به قطر اصلی ماتریس باید نسبت به بقیه خانه‌های ماتریس دارای مقدار بیشتری باشند. شکل شماره 1 را در نظر بگیرید. فرض کنید تصویر (الف) در این شکل، تصویر مورد نظر برای نهان‌نگاری باشد. همچنین فرض کنید ما ماتریس خودهمبستگی را بصورت یک تصویر با ابعاد 256×256 نشان دهیم و مقدار پیکسل‌های این تصویر همان مقادیر ماتریس خودهمبستگی باشند. در اینصورت تصویر (ب) و (ج) در شکل شماره 1 به ترتیب نمایانگر ماتریس خودهمبستگی قبل و بعد از فرآیند درج در تصویر (الف) شکل 1 خواهد بود.

در مقاله [7] سعی شده است پیام در پیکسل‌هایی درج شود که سطوح خاکستری آنها نسبت به هم مستقل باشند. به عبارت دیگر، در این روش سعی می‌شود سطوح خاکستری که نسبت به هم مستقل هستند تعیین و مورد استفاده واقع شوند. در این حالت است که انجام عملیات درج تاثیر چندانی در این ماتریس نخواهد داشت.

مقاله فوق رخ دادن تصادفی سطوح خاکستری در ماتریس خودهمبستگی را بصورت زیر تعریف کرده است: دو رخداد A و B مستقل از هم در نظر گرفته می‌شوند، اگر احتمالات آنها در رابطه P(AB)=P(A)P(B) صدق کند. حال احتمال ترکیبی دو سطح خاکستری به راحتی از ماتریس خودهمبستگی و احتمال سطوح خاکستری بصورت جداگانه براحتی از هیستوگرام تصویر بدست می‌آید. در نهایت قانون درج را بصورت زیر بیان کرده است:

در رابطه فوق T حد آستانه می‌باشد که باتوجه به اندازه بلوک، میزان حساسیت مطلوب، پارامتر r و ظرفیت مورد انتظار تعیین می‌شود. همچنین d را میزان اختلاف موجود در بلوک تعریف کرده و به صورت زیر محاسبه می‌کنیم.

$$d_1 = \sum_{i=1}^{n-1}\sum_{j=1}^{n} |B(i+1,j) - B(i,j)|^r \quad (4)$$

$$d_2 = \sum_{i=1}^{n}\sum_{j=1}^{n-1} |B(i,j+1) - B(i,j)|^r \quad (5)$$

$$d = d_1 + d_2 \quad (6)$$

در رابطه فوق B(i,j) بیانگر سطح خاکستری پیکسل واقع شده در مختصات (i,j) بلوک B می‌باشد. همچنین پارامتر r را پارامتر میزان تاثیر اختلاف سطح خاکستری دو پیکسل مجاور گوییم. مشخص است هر چه این پارامتر بزرگتر باشد، در صورتیکه اختلاف سطح خاکستری دو پیکسل مجاور زیاد شود، مقدار d بزرگتر خواهد شد.

در بعضی موارد تصاویر دارای نویز می‌باشند و ممکن است وجود این نویز منجر شود یک ناحیه مسطح بعنوان ناحیه نویزی در نظر گرفته شود (مثلاً به خاطر یک پیکسل دارای تفاوت زیاد با همسایگان). برای این منظور بهتر است ابتدا فیلتر نرم‌ساز (میانگین) روی تصویر مورد نظر اعمال شود تا به کمک آن مقدار زیادی از تاثیر نویز کاسته شود (شکل شماره 2).

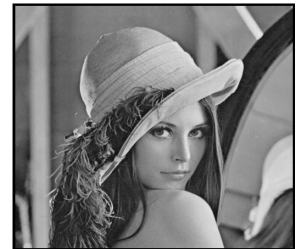

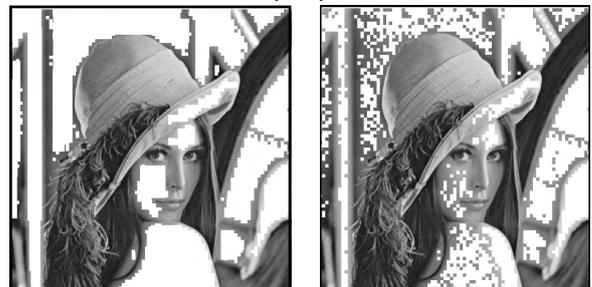

شکل شماره 2 - (الف) تصویر پوشش. (ب) تصویر ناحیه‌بندی شده بدون اعمال فیلتر نرم‌ساز. (ج) تصویر ناحیه‌بندی شده با اعمال فیلتر نرم‌ساز

توجه شود که خروجی این مرحله دو ناحیه کلی خواهد بود که عبارتند از ناحیه نویزی و ناحیه مسطح. در بخش بعد دوباره ناحیه مسطح (فقط ناحیه مسطح) برای درج پیام بلوک‌بندی می‌شود که اندازه بلوک‌های آن می‌تواند از اندازه n متفاوت باشد. در این مقاله برای بلاک‌بندی تصویر مقدار n برابر 6 و مقدار r برابر 4 و مقدار T برابر 2500 در نظر گرفته شده است.

در شکل شماره 2 نواحی که به رنگ سفید نشان داده شده‌اند، بعنوان نواحی مسطح انتخاب گردیده‌اند.

### 2-3- نهان‌نگاری در نواحی نویزی

در نواحی نویزی همبستگی بین مقدار پیکسل‌ها بسیار ناچیز بوده و ماتریس خود همبستگی برای این نواحی حاوی اطلاعات مفیدی برای نهان‌کاوی نمی‌باشد ولی هیستوگرام تصویر در اثر درج پیام دچار تغییراتی می‌شود. برای این منظور برای درج پیام در نواحی نویزی روشی جدید در ادامه ارائه می‌شود که نه تنها در برابر حمله هیستوگرام مقاوم است، بلکه منجر به افزایش 1/5 برابری ظرفیت می‌شود.

با توجه به اینکه در روش پیشنهادی تصویر اصلی در گیرنده موجود است، فرستنده می‌تواند برای درج پیام در هر پیکسل سه حالت در نظر بگیرد: 1) مقدار پیکسل تغییر نکند. 2) مقدار پیکسل یک واحد کاهش یابد. 3) مقدار پیکسل یک واحد افزایش یابد. حال برای اینکه بتوان از هر سه حالت فوق استفاده کرد، ما عملیات درج را همزمان در 2 پیکسل انجام می‌دهیم. مشخص است که در این صورت 9=3×3 حالت بوجود می‌آید که در اینصورت می‌توان سه بیت پیام را در دو پیکسل درج نمود.

نتایج آزمایشات نشان می‌دهد که روش پیشنهادی از لحاظ مقاومت در برابر حمله هیستوگرام مانند روش‌های [7،8،9،10] عمل می‌کند.

### 3-3- نهان‌نگاری در نواحی مسطح

همانطور که در بخش‌های قبل بیان شد، اعمال تغییرات در مقدار پیکسل‌های نواحی مسطح (که همبستگی بین پیکسل‌ها در آن زیاد است) به شدت روی اطلاعات آماری مراتب بالا تاثیر می‌گذارد. در این بخش روش جدیدی برای درج اطلاعات ارائه می‌دهیم که بدون کاهش چشمگیر ظرفیت نهان‌نگاری منجر به کمترین تغییرات آماری مراتب بالا می‌شود. قبل از هر چیز یادآوری می‌کنیم که در روش پیشنهادی، گیرنده، تصویر اصلی را در اختیار دارد. برای این منظور ابتدا نواحی مسطح تصویر را به بلوک‌هایی با اندازه m×m تقسیم می‌کنیم. برای مثال بلوکی از نواحی مسطح تصویر با ابعاد 3×3 زیر را در نظر بگیرید (توجه شود که اندازه بلوک برای بخش مربوط به بلوک‌بندی تصویر برای تعیین بلوک‌های نویزی و مسطح n×n بود، و اندازه بلوک‌هایی که در این بخش استفاده می‌شود m×m است. بنابراین n باید مضربی از m باشد).

| x1 | x2 | x3 |
| x4 | x5 | x6 |
| x7 | x8 | x9 |

در اینصورت بر اساس اینکه مقدار کدامیک از پیکسل‌های فوق تغییر پیدا کند، می‌توان اقدام به درج اطلاعات محرمانه نمود. در ساده‌ترین حالت فرض کنید می‌خواهیم در هر بلوک 3×3 پیامی بطول 3 بیت درج کنیم. در اینصورت بر اساس مقدار پیام مورد نظر، کافی است مقدار پیکسل متناظر آنرا تغییر دهیم. مثال زیر را در نظر بگیرید (مقدار پیام متناظر با تغییر هر پیکسل می‌تواند بر اساس کلید درج و استخراج تعیین شود و ثابت نباشد- مثلاً هر کدام از چهار ماتریس درج زیر می‌تواند برای درج و استخراج پیام بکار رود.).

| 000 | 001 | 010 |
| 011 | 100 | 101 |
| 110 | 111 | --- |

| 011 | 000 | 010 |
| 111 | --- | 001 |
| 110 | 100 | 101 |

| 000 | 110 | 101 |
| 011 | 001 | 010 |
| --- | 111 | 100 |

| --- | 001 | 010 |
| 011 | 101 | 100 |
| 000 | 111 | 110 |

برای مثال فرض کنید می‌خواهیم پیام (001101100010) را در ناحیه مسطح تصویر زیر (چهار عدد بلوک) بر اساس ماتریس درج A، درج نماییم. ماتریس‌های زیر نتیجه را نشان می‌دهند.

| 000 | 110 | 101 |
| 011 | 001 | 010 |
| --- | 111 | 100 |

ماتریس درج A

| 78 | 78 | 79 |   | 80 | 82 | 83 |
| 78 | 79 | 79 |   | 81 | 82 | 84 |
| 79 | 81 | 82 |   | 81 | 83 | 85 |

| 80 | 83 | 83 |   | 83 | 84 | 85 |
| 82 | 83 | 84 |   | 85 | 86 | 85 |
| 83 | 84 | 85 |   | 86 | 88 | 88 |

ماتریس تصویر قبل از درج پیام محرمانه

| 78 | 78 | 79 |   | 80 | 82 | 82 |
| 78 | 78 | 79 |   | 81 | 82 | 84 |
| 79 | 81 | 82 |   | 81 | 83 | 85 |

| 80 | 83 | 83 |   | 83 | 84 | 85 |
| 82 | 83 | 84 |   | 85 | 86 | 86 |
| 83 | 84 | 84 |   | 86 | 88 | 88 |

ماتریس تصویر بعد از درج پیام محرمانه

برای درج پیام به‌روش فوق در نظر گرفتن نکات زیر ضروری است:

1) می‌توان بر اساس طول پیامی که باید در هر بلوک درج شود، سیاست‌های دیگری هم در نظر گرفت. برای سیاست فوق با در نظر گرفتن اینکه مقدار پیکسل مورد نظر یک واحد کاهش یا افزایش داشته باشد، می‌توان در هر بلوک 4 بیت پیام را درج نمود (که علاوه بر افزایش ظرفیت، هیستوگرام تصویر هم حفظ شود). در صورتی که نخواهیم از این تکنیک برای افزایش ظرفیت نهان‌نگاری استفاده کنیم، باید آنرا برای حفظ هیستوگرام تصویر لحاظ نماییم.

2) برای اینکه انجام عمل درج منجر به کمترین خرابی در همبستگی بین پیکسل‌های مجاور شود، بهتر است قبل از تغییر پیکسل متناظر، همسایه‌های آن را در نظر گرفته، پیکسل مورد نظر بر اساس مقدار آنها تغییر یابد.

3) مثال ذکر شده برای بلوک‌های با اندازه m=3 بیان شد. در واقع بر اساس طول پیام و سیاست درج (کلید) هر اندازه‌ای را می‌توان برای m در نظر گرفت (البته با اعمال محدودیت m≤n).

4) محدودیتی برای انتخاب ماتریس مرجع وجود ندارد و این ماتریس می‌تواند بر اساس کلید درج انتخاب شود.

5) با توجه به طبیعت نویزی بودن تصاویر، نتایج آزمایشات نشان می‌دهد که اعمال فیلتر نرم‌ساز قبل از ناحیه‌بندی تصویر منجر به نتایج بهتری می‌شود.

## 4- آزمایشات

در آزمایش اول قصد داریم روش خود را از لحاظ خرابی ادراکی بوجود آمده در نتیجه درج پیام با اندازه 0.8bpp با تعدادی از روش‌های نهان‌نگاری مقایسه کنیم. برای این منظور پیام با اندازه مذکور را در 1000 عدد تصویر با ابعاد 500×500 که از لحاظ بافت، محتوی و کانتراست متفاوت بودند درج نمودیم. شکل شماره 3 تعدادی از این تصاویر را نشان می‌دهد. نتیجه این آزمایش نشان داد که معیار PSNR برای چهار روش درج LSB ، $LSB^+$ ، LSB±1 و روش پیشنهادی بطور متوسط برابر 51/1 ، 49/3، 47/1 و 53/7 می‌باشد. این نتیجه بیانگر این مطلب است که روش پیشنهادی نسبت به سه روش دیگر دارای کمترین خرابی ادراکی می‌باشد.

ادراکی و آماری کمتری نسبت به روش‌های رایج نهان‌نگاری می‌شود. برای این منظور ابتدا تصویر به ناحیه‌های نویزی و مسطح تقسیم می‌کنیم و برای درج پیام در هر ناحیه روش خاصی پیشنهاد شده است. نتایج آزمایشات بیانگر این مطلب است که روش ما علاوه بر افزایش ظرفیت، در مقایسه با تعدادی از روش‌های نهان‌نگاری در حوزه مکان، منجر به خرابی ادراکی کمتری شده و حملات هیستوگرام و آماری مراتب بالا با قدرت خیلی کمتر قادر به تشخیص آن می‌باشند.

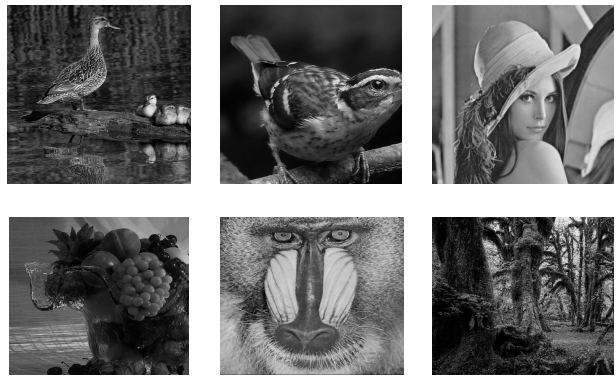

شکل 3- تعدادی از تصاویر نمونه مورد استفاده برای آزمایشات

جدول 1: مقایسه روش پیشنهادی از نظر امنیت (TP)

| نام حمله | روش LSB | روش $LSB^+$ | روش LSB±1 | روش پیشنهادی |
|---|---|---|---|---|
| [5] | 94% | 0.07% | 0.084% | 0.081% |
| [6] | 96.1% | 97.5% | 0.17% | 0.08 |
| [12] | 97% | 98.2% | 98% | 0.12% |
| [13] | 99% | 87.3% | 0.09% | 0.05% |

در آزمایش دوم روش پیشنهادی را از لحاظ امنیت با تعدادی از روش‌های نهان‌نگاری مقایسه می‌کنیم. برای این منظور پیام با اندازه 0.5bpp را در 1000 تصویر مذکور درج کرده و سپس آنها را با تعدادی از روش‌های نهان‌کاوی مورد بررسی قرار دادیم. جدول شماره 2 نتیجه بدست آمده را نشان می‌دهد.

بر اساس جدول شماره 2 مشخص می‌شود روش LSB در مقابل تمامی حملات از جمله حمله هیستوگرام[5] با شکست مواجه شده است، ولی با توجه به اینکه سایر روش‌ها از مکانیزم مقابله با این حمله برخوردار بوده‌اند، حمله هیستوگرام با شکست مواجه شده است. همچنین نتایج حاصل از نهان‌کاوی به کمک سه روش دیگر بیانگر آن است که روش پیشنهادی نسبت به روش‌های دیگر در مقابل حملات آماری مراتب بالا شکست ناپذیر است. این سه روش همبستگی بین پیکسل‌ها را در نظر گرفته و بر اساس این همبستگی اطلاعات آماری مراتب بالا را استخراج می‌کنند. حمله مرجع [6] معروف به حمله POV می‌باشد. روش مربوط به مرجع [11] از ماتریس خودهمبستگی استفاده می‌نماید. و در نهایت روش [13] تحت عنوان حمله RS شناخته می‌شود. همانطور که جدول شماره 2 نشان می‌دهد، روش پیشنهادی در مقابل این سه حمله مقاوم بوده است.

## 5- نتیجه‌گیری

در این مقاله روش تطبیقی جدیدی برای نهان‌نگاری در تصاویر ارائه شد که منجر به خرابی‌های